\documentclass[]{revtex4}
\usepackage{hyperref}
 
\begin{document}

\title{The quantum description of BF model in superspace}

 \author{ Manoj Kumar Dwivedi}
 \email {manojdwivedi84@gmail.com} 
 
\affiliation { Department of Physics, Institute of Science, Banaras Hindu University,  \\
Varanasi-221005, India.}

\begin{abstract}
We consider the BRST symmetric four dimensional BF theory, a topological theory, containing antysymmetric 
tensor fields in Landau gauge and  extend the BRST symmetry by introducing 
a shift symmetry to it. Within this formulation, the antighost fields corresponding to 
shift symmetry coincide  with antifields of standard field/antifield formulation.
Further, we provide a superspace description for the BF model possessing extended BRST 
and extended anti-BRST transformations.
 \end{abstract}
\maketitle

\section{  Introduction}
Topological gauge field theories (TGFT) which came from mathematics have some peculiar features. 
The examples of two distinct class of TGFT are  topological Yang-Mills theory and Chern-Simons (CS) theory, 
which are some times classified as Witten-type and of Schwarz-type respectively \cite{BF1}. 
Except these two types, there are another Schwarz-type TGFT called topological BF theory, 
which is an extension of CS theory \cite{BF2}. The difference between CS theory and BF model is 
that action of previous theory exist only in odd-dimensions while later one can be defined on 
manifolds of any dimensions.

In string theory and non-linear sigma model, four dimensional antisymmetric (or BF) model \cite{BF3} were introduced some years ago. 
This model is interested due to its topological nature \cite{BF1} and 
 their connection with lower dimensional quantum gravity,  for example three space-time dimensional Einstein-Hilbert 
with or without using cosmological constant can be naturally formulated in terms of BF-models\cite{BF4,BF5}. 
Coupling of an antisymmetric tensor field with the field strength tensor of Yang-Mills is describe by these
 models \cite{BF6}. Quantization of BF model in Landau gauge has been studied in Ref. \cite{BF6}. 
Topological BF theory in Landau gauge has a common feature of a large class of
 topological models \cite{BF7, BF8}.

On the other hand, the Batalin-Vilkovisky (BV) approach, also known as field/antifield formulation,
 \cite{BF10,BF11,BF12,BF13} is one of the most powerful quantization algorithms presently available.
 BV formulation   deals with very general gauge theories, including those with open or reducible gauge symmetry algebras.
The BV method also address the possible violations of symmetries of the action by quantum effects.
The BV formulation (independently 
introduced by Zinn-Justin \cite{BF14}) extends the BRST approach \cite{sud1}. 
In fact, the BRST symmetry \cite{BRS,BRS1} is a very important symmetry for gauge theories
\cite{sudup}.
 Beside the covariant description to perform the gauge-fixing
in quantum field theory, BV formulation was also applies to other problems like analysing
possible deformations of the action and anomalies.

A superspace description for various  gauge theories in BV formulation has been studied extensively \cite{bono, ad, ba,sudp, sudb}. 
They have   shown  that the extended BRST and extended
anti-BRST invariant actions of these theories
(including some shift symmetry) in   BV  formulation  yield 
naturally  the proper identification of the antifields through equations of motion.
The shift symmetry is important  and gets relevance, for example,  in  inflation 
particularly in supergravity  \cite{bra} as well as in Standard Model \cite{he}. 
In usual BV formulation, these  antifields  can be calculated from the expression of gauge-fixing fermion.  
We extended BRST formulation and superspace description of  the topological gauge (BF) model is still unstudied and we try to discuss these here.

In the present work, we try to generalize the superspace formulation of BV action for BF model. 
Particularly, we first consider BRST invariant 
BF model in Landau gauge and extend the BRST symmetry of the theory by 
including shift symmetry. By doing so, we find that the antighosts of shift symmetry  get
 identified as antifields of standard BV formulation naturally. 
 Further, we discuss a superspace formulation of  extended BRST invariant BF model.
 Here we see that  one additional Grassmann coordinate is required if action admits
 only extended BRST symmetry. However, for both extended BRST  and extended 
anti-BRST invariant BF model two additional Grassmann coordinates are required.

This paper is framed as follows. In section II, we discuss the BRST invariant BF model.
In section III, we study the extended BRST transformation of the model.
Further, we describe extended BRST invariant action in superspace in section IV.
The extended anti-BRST symmetry is discussed in section V.
The superspace formulation of extended BRST and anti-BRST invariant action is 
given in section VI.
The last section is reserved for concluding remarks.

\section{BRST invariant BF model}
In this section, we discuss the preliminaries of BF model with
its BRST invariance. In this view, 
the BF model in  flat $(3+1)$  space-time dimensions
is given by the following gauge invariant Lagrangian density \cite{BF6}:
\begin{eqnarray}
{\cal L}_{0}&=&  -\frac{1}{4} \epsilon^{\mu \nu \rho \sigma} F^a_{\mu \nu} B^a_{\rho \sigma},
\end{eqnarray}
where $B^a_{\rho \sigma}$ and $F^a_{\mu \nu}$ are two-form field and field-strength tensor for vector 
field respectively. 
In order to remove discrepancy due to gauge symmetry, the  gauge fixing and ghost terms are given by
\begin{eqnarray}
{\cal L}_{gf + gh}&=&      b^a \partial^\mu A^a_\mu + 
\bar C^a \partial^\mu (D_\mu C)^a + h^{a\nu} (\partial^\mu B^a_{\mu \nu} ) + \omega^a \partial^\mu \xi_\mu^a + 
h^a_\mu (\partial^\mu e^a) \nonumber\\
 &+& \omega^a \lambda^a + (\partial^\mu \bar \xi^a_\mu ) \lambda^a - 
(\partial^\mu \bar \phi^a ) [ (D_\mu \phi )^a + f^{abc} c^b \xi^c_\mu ] \nonumber\\ 
&-& (\partial^\mu \bar \xi^{a \nu} ) [(D_\mu \xi_\nu )^a - (D_\nu \xi_\mu )^a + 
f^{abc} B^b_{\mu \nu} C^c ] \nonumber\\ 
&+& \frac{1}{2} f^{abc} \epsilon^{\mu \nu \rho \sigma} (\partial_\mu \bar \xi^a_\nu ) (\partial_\rho \bar \xi^b_\sigma ) \phi^c ,
\end{eqnarray}
where fields $(C^a,\xi^a_\mu)$, $(\bar C^a, \bar {\xi}^a_\mu)$ and  $(b^a, h^a_\mu)$ are the ghosts,
  antighosts and the multipliers fields respectively, while the fields $\phi^a,\bar{\phi}^a$ and $\omega^a$ are 
taken into account to remove further degeneracy due to the existence of zero modes in the transformations.

The effective Lagrangian density  
of BF model, ${\cal L} ={\cal L}_{0}+{\cal L}_{gf + gh}$,
  possesses following BRST symmetry:
\begin{eqnarray}
&&s A^a_\mu = - ( D_\mu C )^a,  \ \
 s C^a  = \frac{1}{2} f^{abc} C^b C^c,  \ \
 s \xi^a_\mu = (D_\mu \phi)^a + f^{abc} C^b \xi^c_\mu, \nonumber\\
&&s B^a_{\mu \nu} = -( D_\mu \xi_\nu - D_\nu \xi_\mu )^a - f^{abc} B^b_{\mu \nu} C^c + 
f^{abc} \epsilon_{\mu \nu \rho \sigma} (\partial^\rho \bar \xi^{b \sigma}) \phi^c \nonumber\\
&&s \phi^a = f^{abc} C^b \phi^c, \ \  s \bar \xi^a_\mu = h^a_\mu, \ \
 s \bar C^a = b^a, \ \
 s \bar \phi^a = \omega^a,  \ \
 s e^a = \lambda^a,  \nonumber\\
&&s (h^a_\mu, b^a, \omega^a, \lambda^a ) = 0 .     
\end{eqnarray}
The  gauge-fixing and ghost terms of the effective Lagrangian density  is
BRST exact and, hence, can be written in terms of BRST variation of
gauge-fixing fermion,  
\begin{equation}
\Psi = \left(\bar C^a \partial^\mu A^a_\mu + \bar \xi^{a\mu} \partial^\nu B^a_{\mu \nu} + 
\bar \phi^a \partial^\mu \xi^a_\mu - e^a \omega^a - e^a \partial^\mu \bar \xi^a_\mu \right),
\label{ps1}
\end{equation} 
 as follows
\begin{equation}
{\cal L}_{gf + gh} =   s\Psi.
\label{ps}
\end{equation}
In the next section, we would like to study the extended BRST symmetry  for the model which incorporates shift symmetry together with original BRST symmetry.
\section{Extended BRST Invariant Lagrangian Density}
 The advantage of studying the extended BRST transformations for BF model
in BV formulation   is that antifields 
get identification naturally. We begin with shifting  all the fields
from their original value as follows,
\begin{eqnarray}
&&B^a_{\mu \nu} \longrightarrow  B^a_{\mu \nu} - \tilde B^a_{\mu \nu} ,\quad
A^a_\mu \longrightarrow  A^a_\mu - \tilde A^a_\mu, \quad
C^a \longrightarrow  C^a - \tilde C^a,\nonumber\\
&&
\bar C^a \longrightarrow \bar C^a - \tilde {\bar C}^a, \quad
b^a \longrightarrow b^a - \tilde b^a, \quad
\xi^a_\mu \longrightarrow  \xi^a_\mu - \tilde \xi^a_\mu,\nonumber\\
 &&
\bar \xi^a_\mu \longrightarrow  \bar \xi^a_\mu - \tilde {\bar \xi^a_\mu}, \quad
\phi^a \longrightarrow  \phi^a - \tilde \phi^a, \quad
\bar \phi^a \longrightarrow  \bar \phi^a - \tilde {\bar\phi^a}, \quad
h^a_\mu \longrightarrow  h^a_\mu - \tilde h^a_\mu,\nonumber\\
&& e^a \longrightarrow  e^a - \tilde e^a, \quad
\omega^a \longrightarrow  \omega^a - \tilde \omega^a, \quad
\lambda^a \longrightarrow  \lambda^a - \tilde \lambda^a.
\end{eqnarray}
The effective Lagrangian density of BF model also get shifted under such shifting of 
fields respectively. This is given by
\begin{eqnarray}
\tilde{\cal L } &=& {\cal L} (A^a_\mu - \tilde A^a_\mu, C^a - \tilde C^a, \bar C^a - 
\tilde {\bar C^a}, b^a - \tilde b^a, \xi^a_\mu - \tilde \xi^a_\mu, 
 \bar \xi^a_\mu - \tilde {\bar \xi^a_\mu},\nonumber\\
&& \phi^a - \tilde \phi^a, \bar \phi^a - 
\tilde {\bar\phi^a}, h^a_\mu - \tilde h^a_\mu , e^a - \tilde e^a, 
\omega^a - \tilde \omega^a, \lambda^a - \tilde \lambda^a ).\label{shif}
\end{eqnarray}
The shifted Lagrangian density is invariant 
under BRST transformation together with a shift symmetry transformation, jointly
known as extended BRST transformation.
The extended BRST 
symmetry transformations under which Lagrangian density of BF model is invariant are
written by
\begin{eqnarray}
s A^a_\mu &=&\psi^a_\mu ,\ \
s \tilde A^a_\mu =\psi^a_\mu - ( D_\mu - \tilde D_\mu ) (C - \tilde C)^a, \ \
s C^a = \epsilon^a  ,\ \
s \tilde C^a = \epsilon^a - \frac{1}{2} f^{a b c} (C^b-\tilde{C}^b) (C^c -\tilde C^c),  \nonumber\\
s \tilde {\bar C^a} &=& \bar \epsilon^a - (b -\tilde b)^a,\ \
s b^a = \chi^a ,\ \
s \tilde b^a  =\chi^a, \ \
s \phi^a = M^a ,\ \
s \tilde \phi^a = M^a - f^{a b c} (C^b - \tilde{C}^b )(\phi^c - \tilde{\phi}^c),\nonumber\\
s \bar \phi^a &=& \bar M^a ,\ \
s \tilde {\bar\phi^a} = M^a - (\omega^a - \tilde {\omega}^a),\ \
s e^a = N^a ,\ \
s \tilde e^a = N^a - (\lambda^a - \tilde {\lambda}^a),\ \
s \xi^a_\mu = L^a_\mu , \ \
s \bar C^a = \bar \epsilon^a ,\nonumber\\
s \tilde \xi^a_\mu &=& L^a_\mu - [ (D_\mu - \tilde D_\mu ) (\phi - \tilde \phi )^a + f^{abc} (C^b  - \tilde C^b) (\xi^c_\mu - \tilde\xi^c_\mu ) ],\ \
s \bar \xi^a_\mu = \bar L^a_\mu ,\ \ 
s \tilde {\bar\xi^a_\mu} = \bar L^a_\mu - (h^a_\mu - \tilde{h}^a_\mu ), \label{ex}
\end{eqnarray}
where $\psi^a_\mu, \epsilon^a, \bar \epsilon^a, \chi^a, M^a,\bar M^a, N^a, L^a_\mu,$ 
and $\bar L^a_\mu$ are the ghost fields corresponding to shift
symmetry for $A^a_\mu , C^a ,\bar C^a, b^a, \phi^a, \bar \phi^a, e^a, \xi^a_\mu$ 
and $\bar \xi^a_\mu$ respectively.
The nilpotency of extended BRST symmetry (\ref{ex}) leads to 
the BRST transformation for the  following  ghost fields:
\begin{eqnarray}
s \psi^a_\mu &=& 0 , \ \
s \epsilon^a = 0 , \ \
s \bar \epsilon^a = 0 , \ \
s \chi^a  = 0 , \ \
s M^a  = 0 , \nonumber\\
s \bar M^a  &=& 0 , \ \
s N^a  = 0 , \ \
s L^a_\mu = 0 , \ \
s \bar L^a_\mu = 0.
\end{eqnarray}
In order to make the theory  ghost free,  we need 
further   antighosts $A^{\star a}_\mu, C^{\star a}, \bar 
C^{\star a}, b^{\star a}, \xi^{\star a}_\mu, \bar \xi^{\star a}_\mu, \phi^{\star a}, 
\bar \phi^{\star a}, $ and $e^{\star a}$ to be introduced corresponding to the ghost fields
 $\psi^a_\mu, \epsilon^a, \bar \epsilon^a, \chi^a, M^a,\bar M^a, N^a, L^a_\mu,$ 
and $\bar L^a_\mu$ respectively. The  
 BRST transformations of these antighosts are constructed as follows
\begin{eqnarray}
s A^{\star a}_\mu &=& -\zeta^a_\mu , \ \
s C^{\star a} = - \sigma^a, \ \
s \bar C^{\star a} = -\bar \sigma^a , \ \
s b^{\star a} = - \varpi^a, \ \
s \phi^{\star a} = - \upsilon^a, \nonumber\\
s \bar \phi^{\star a} &=& -\bar \upsilon^a, \ \
s e^{\star a} = -\tau^a, \ \
s \xi^{\star a}_\mu = -\kappa_\mu^a, \ \
s \bar \xi^{\star a}_\mu = -\bar \kappa_\mu^a, \ \
\end{eqnarray} 
where $\zeta^a_\mu, \sigma^a , \bar \sigma^a, \varpi^a, \upsilon^a, \tau^a, \kappa_\mu^a,  $ and $\bar \kappa_\mu^a$ are
the  Nakanishi-Lautrup type auxiliary fields 
corresponding to shifted fields $\tilde A^{ a}_\mu, \tilde  C^{a}, \tilde {\bar C^{a}},
\tilde  b^{ a}, \tilde \phi^{a}, \tilde {\bar \phi}^{a},
\tilde e^{a}, \tilde \xi^{a}_\mu,  $ and $\tilde {\bar \xi^{a}_\mu}$ having following
 BRST transformations:
\begin{eqnarray}
s \zeta^a_\mu &=& 0 , \ \
s \sigma^a  = 0 , \ \
s \bar \sigma^a  = 0 , \ \
s \varpi^a  = 0 , \ \
s \upsilon^a  = 0 , \nonumber\\
s \bar \upsilon^a &=& 0 , \ \
s \tau^a = 0 , \ \
s \kappa_\mu^a  = 0 , \ \
s \bar \kappa_\mu^a = 0.
\end{eqnarray}
We can recover our original BF model by  fixing the shift symmetry  in such a way such that 
effect of all the tilde fields will vanish.
We achieve this by adding following gauge-fixed term
to the shifted Lagrangian density (\ref{shif}):
\begin{eqnarray}
\tilde{\cal L}_{gf+gh} &=&  -\zeta^{a \mu} \tilde A^a_\mu - A_\mu^{a \star} [\psi^{a \mu} - 
(D^\mu - \tilde D^\mu ) (C - \tilde C)^a] - \bar \sigma^a  \tilde  C^a 
+ \bar C^{a \star}[\epsilon^a  - \frac{1}{2} f^{a b c}(C^b-\tilde{C}^b) (C^c -\tilde C^c)] \nonumber\\
& -&   \sigma^a  \tilde {\bar C^a} +  C^{a\star}[\bar \epsilon^a  - (b^a - \tilde b^a)] -  \upsilon^a 
\tilde \phi^a - \phi^{\star a} \left[ M^a - f^{a b c} (C^b - \tilde{C}^b )(\phi^c - \tilde{\phi}^c) \right] - \bar \upsilon^a \tilde {\bar \phi^a}
 \nonumber\\ &-&  \bar\phi^{\star a} (\bar M^a - \omega^a +\tilde \omega^a) - \tau^a \tilde e^a  -e^\star[N^a - (\lambda^a -\tilde\lambda^a)] 
- \varpi^a \tilde b^a - b^{\star a} \chi^a - \kappa^{\mu a} \tilde \xi^a_\mu - 
\bar \kappa^{\mu a} \tilde {\bar \xi^a_\mu} \nonumber\\ 
&+&  
\xi^{\star \mu a} \left(L^a_\mu - \left[ (D_\mu - \tilde D_\mu ) (\phi - \tilde \phi )^a + f^{abc} (C^b  - \tilde C^b) (\xi^c_\mu - \tilde\xi^c_\mu ) \right] \right)  + \bar \xi^{\star \mu a} \left[\bar L^a_\mu - (h^a_\mu - \tilde{h}^a_\mu )\right].\label{i}
\end{eqnarray}
One can easily check that this gauge-fixing Lagrangian density $\tilde{\cal L}_{gf+gh}$  also 
admits the extended BRST invariance.
Integrating the auxiliary fields of the
above expression, we obtain
\begin{eqnarray}
\tilde{\cal L}_{gf+gh} &=&  - A_\mu^{a \star} [\psi^{a\mu} - (D^\mu  C)^a] 
+  \bar C^{a\star}[\epsilon^a  - \frac{1}{2} f^{a b c}  C^b C^c   ] \nonumber\\
& +&   C^{a\star}[\bar \epsilon^a  -  b^a  ] -  
 \phi^{\star a} ( M^a - f^{a b c}  C^b  \phi^c   ) 
 \nonumber\\ &-&  \bar\phi^{\star a} (\bar M^a - \omega^a)  - e^{\star a} [N^a - \lambda^a ] 
- b^{\star a} \chi^a \nonumber\\ &+&  
\xi^{\star \mu a} \left(L^a_\mu - [  (D_\mu  \phi)^a + f^{abc}  C^b  \xi^c_\mu    ] \right) 
 + \bar \xi^{\star \mu a} [\bar L^a_\mu - h^a_\mu ]. 
\end{eqnarray}
The gauge-fixing and ghost terms of the Lagrangian density 
are BRST exact and can be expressed   in terms of a general gauge-fixing fermion 
$\Psi$ as 
\begin{eqnarray}
  s \Psi &=&  s A^a_\mu \frac{\delta \Psi}{\delta A^a_\mu} + s C^a 
 \frac{\delta \Psi}{\delta C^a} 
+ s \bar C^a \frac{\delta \Psi}{\delta \bar C^a} +  s b^a \frac{\delta \Psi}{\delta b^a}
 +  s \xi^a_\mu \frac{\delta \Psi}{\delta\xi^a_\mu} 
+  s \bar \xi^a_\mu \frac{\delta \Psi}{\delta\bar \xi^a_\mu} 
+ s  \phi^a \frac{\delta \Psi}{\delta \phi^a} + s \bar \phi^a \frac{\delta \Psi}{\delta \bar \phi^a}
+ s  e^a \frac{\delta \Psi}{\delta e^a}, 
\nonumber\\
&= & - \frac{\delta \Psi}{\delta A^a_\mu}\psi^a _\mu + \frac{\delta \Psi}{\delta C^a} 
\epsilon^a
+ \frac{\delta \Psi}{\delta \bar C^a}\bar\epsilon^a - \frac{\delta \Psi}{\delta b^a} \chi^a
- \frac{\delta \Psi}{\delta \xi^a_\mu} L^a_\mu - \frac{\delta \Psi}{\delta \bar \xi^a_\mu} \bar L^a_\mu
- \frac{\delta \Psi}{\delta \phi^a} M^a - \frac{\delta \Psi}{\delta \bar \phi^a} \bar M^a - 
\frac{\delta \Psi}{\delta e^a} N^a.
\label{g}
\end{eqnarray} 
After integrating out the auxiliary fields which set the tilde fields to zero, we have
  the complete effective action for BF model in landau gauge possessing extended
  BRST symmetry  as
\begin{eqnarray}
{\cal L}_{eff} & = &   {\cal L}_0 + {\cal L}_{gf + gh} 
+\tilde{\cal L}_{gf+gh},\nonumber\\
  &=&{\cal L}_0  +   \left(- A_\mu^{\star a} - \frac{\delta \Psi}{\delta A^{\mu a}}
  \right)\psi ^{\mu a} +
  \left(\bar C^{\star a} + \frac{\delta \Psi}{\delta C^a}\right) \epsilon^a +
  \left(C^{\star a} + \frac{\delta \Psi}{\delta \bar C^a}\right)\bar\epsilon^a 
- \left(b^{\star a} + \frac{\delta \Psi}{\delta b^a}\right) \chi^a  \nonumber\\ 
 &+&  \left(\xi^{\star a}_\mu + \frac{\delta \Psi}{\delta \xi^a_\mu}\right)L^{\mu a} 
 + \left(\bar \xi^{\star a}_\mu + \frac{\delta \Psi}{\delta \bar \xi^a_\mu}\right)\bar L^{\mu a} 
 - \left(\phi^{\star a} + \frac{\delta \Psi}{\delta \phi^a}\right)M^a 
- \left(\bar \phi^{\star a} + \frac{\delta \Psi}{\delta \bar \phi^a}\right)\bar M^a   \nonumber\\  
 &+&   \left(- e^{\star a} - \frac{\delta \Psi}{\delta e^a}\right) N^a
  + A_\mu^{a\star} [(D^\mu  C^a) - \frac{\bar C^{a\star}}{2} f^{a b c}  C^b \xi^{\mu c}] 
+ C^{\star a} b^a \nonumber\\  
&+& \xi^{\star \mu a}[ (D_\mu  \phi)^a + f^{abc}  C^b  \xi^c_\mu  
+   \phi^{\star a} f^{abc} C^b \xi^c_\mu].\label{eff}
\end{eqnarray}
Integrating out the  ghost fields associated with shift symmetry, we obtain
\begin{eqnarray}
A_\mu ^{a \star} &=& -\frac{\delta \Psi}{\delta A^{\mu a}} , \ \
\bar C^{\star a} = - \frac{\delta \Psi}{\delta C^a}  , \ \
C^{\star a} = - \frac{\delta \Psi}{\delta \bar C^a}  , \ \
b^{\star a} = -\frac{\delta \Psi}{\delta b^a}   , \nonumber\\
\xi^{\star a}_\mu &=& -\frac{\delta \Psi}{\delta \xi^{\star a}_\mu} , \ \
\bar \xi^{\star a}_\mu = -\frac{\delta \Psi}{\delta \bar \xi^{\star a}_\mu} , \ \
\phi^{\star a} =- \frac{\delta \Psi}{\delta \phi^a} , \ \
\bar \phi^{\star a} =-\frac{\delta \Psi}{\delta \bar \phi^a},\ \ 
e^{\star a} = -\frac{\delta \Psi}{\delta e^a}.  
\end{eqnarray}
For a particular choice of gauge-fixing fermion $\Psi$ given in (\ref{ps1}), anti-ghost fields get following identifications:
\begin{eqnarray}
A_\mu ^{a\star} &=&  \partial_\mu \bar {C}^a, \ \
\bar C^{a\star} = 0, \ \
C^{a\star} = -\partial_\mu A^{\mu a}, \ \
b^{a\star} = 0 , \ \
\xi^{\star a}_\mu =    \partial_\mu\bar \phi^a, \nonumber\\
\bar \xi^{\star a}_\mu &=&   - \partial^\nu B^a_{\mu\nu}-\partial_\mu e^a, \ \
\phi^{\star a} =  0, \ \
\bar \phi^{\star a} =- \partial^\mu \xi_\mu^a, \ \
e^{\star a} = \omega^a+\partial^\mu\bar\xi^a_\mu.
\end{eqnarray}
It is obvious to see that with these anti-ghost fields, the expression (\ref{eff}) changes to 
the original Lagrangian density of  the BF model in Landau gauge.

\section{Extended BRST invariant superspace description}
In this section, the Lagrangian density of BF model which is 
invariant under the extended BRST transformations only is described 
in a superspace $(x_\mu, \theta)$, where $\theta$ is a Grassmann coordinate and $x_\mu$ is the four dimanesional spect-time coordinates.
In order to give superspace description for the extended BRST invariant theory, we first define superfields of the form:
\begin{eqnarray}
A^a_\mu (x,\theta ) &=& A^a_\mu + \theta \psi^a_ \mu , \ \
\tilde A^a_\mu (x,\theta ) = \tilde A^a_\mu + \theta [\psi_ \mu - (D_\mu - \tilde D_\mu )(C - \tilde C)]^a , 
\nonumber\\
\chi^a (x,\theta ) &=& C^a + \theta \epsilon^a , \ \
\tilde \chi^a (x,\theta )= \tilde C^a + \theta [\epsilon^a - \frac{1}{2} f^{a b c} (C^b  - \tilde C^b) (C^c - \tilde C^c )] , \nonumber\\
\bar {\chi}^a (x,\theta ) &=& \bar C^a + \theta \bar \epsilon^a , \ \
\tilde {\bar \chi^a} (x,\theta )= \tilde {\bar C^a} + \theta [\bar \epsilon^a - (b - \tilde b)^a ] , \nonumber\\
b^a (x,\theta ) &=& b^a + \theta \chi^a , \ \
\tilde b^a (x,\theta ) = \tilde b^a + \theta \chi^a , \ \
\xi^a_\mu (x,\theta ) = \xi^a_\mu + \theta L^a_ \mu , \nonumber\\
\tilde \xi^a_\mu (x,\theta ) &=& \tilde \xi^a_\mu + 
\theta [ L^a_ \mu - [ (D_\mu - \tilde D_\mu ) (\phi - \tilde \phi )^a + f^{abc} (C^b  - \tilde C^b) (\xi^c_\mu - \tilde\xi^c_\mu ) ] ]  , \ \
\bar \xi^a_\mu (x,\theta ) =\bar \xi^a_\mu + \theta \bar L^a_ \mu , \nonumber\\
\tilde {\bar \xi^a_\mu} (x,\theta ) &=& \tilde {\bar \xi^a_\mu} + \theta \left[\bar L^a_\mu - (h^a_\mu - \tilde{h}^a_\mu )\right], \ \
\phi^a (x,\theta ) = \phi^a + \theta M^a , \nonumber\\
\tilde \phi^a (x,\theta ) &=& \tilde \phi^a + \theta [ M^a - f^{a b c} (C^b - \tilde C^b)( \phi^c - \tilde \phi^c )] , \ \
\bar \phi^a (x,\theta ) = \bar \phi^a + \theta \bar M^a , \nonumber\\
\tilde{\bar \phi^a} (x,\theta ) &=& \tilde{\bar \phi^a} + \theta [\bar M^a - (\omega^a-\tilde \omega^a)] , \ \
e^a (x,\theta )= e^a + \theta N^a ,  \
\tilde e^a (x,\theta ) = \tilde e^a + \theta [N^a - (\lambda^a - \tilde \lambda^a)].
\end{eqnarray}
The super-antifields in superspace are defined as follows 
\begin{eqnarray}
\tilde A^{\star a}_\mu (x,\theta ) &=&   A^{\star a}_\mu - \theta \zeta^a_\mu , \ \
\tilde \chi^{\star a} (x,\theta )=   C^{\star a} - \theta  \sigma^a , \ \
\tilde {\bar \chi}^{\star a} (x,\theta ) =   {\bar C^{\star a}} - \theta \bar \sigma^a , \nonumber\\
\tilde b^{\star a} (x,\theta ) &=&   b^{\star a} - \theta \varpi^a , \ \
\tilde \xi^{\star a}_\mu (x,\theta ) =   \xi^{\star a}_\mu - \theta \kappa_\mu^a  , \ \
\tilde {\bar \xi^{\star a}_\mu} (x,\theta ) =   {\bar \xi^{\star a}_\mu} - \theta\bar \kappa_\mu^a , \nonumber\\
\tilde \phi^{\star a} (x,\theta ) &=&  \phi^{\star a} - \theta \upsilon^a , \ \
\tilde{\bar \phi^{\star a}} (x,\theta ) =  {\bar \phi^{\star a}} - \theta \bar \upsilon^a , \ \
\tilde e^{\star a} (x,\theta ) =   e^{\star a} - \theta \tau^a.
\end{eqnarray}
From the above expressions of superfields and super-antifields, we calculate 
\begin{eqnarray}
\frac{\delta(\tilde A_\mu^{a\star} \tilde A^{a\mu})}{\delta \theta} &=& 
-A_\mu^{a\star}[\psi^{a\mu} - (D^\mu - \tilde D^\mu ) (C - \tilde C)^a] - \zeta^a_\mu \tilde A^{a \mu} , \nonumber\\
\frac{\delta (\tilde {\bar\chi} ^{a\star} \tilde \chi^a)}{\delta \theta} &=& 
\bar C^{a\star}[\epsilon^a - \frac{1}{2} f^{a b c} (C^b - \tilde C^b) (C^c - \tilde C^c ) ] - \bar\sigma^a  \tilde C^a , \nonumber\\
\frac{\delta (\tilde {\bar \chi^a} \tilde \chi^{a\star})}{\delta \theta} &=& -  \sigma^a \tilde {\bar 
C^a}  +   C^{a\star}[ \bar \epsilon^a - (b^a - \tilde b^a )], \nonumber\\
\frac{\delta (\tilde b^{a\star} \tilde b^a)}{\delta \theta} &=&-b^{a\star} \chi^a - \varpi^a \tilde b^a, \nonumber\\
\frac{\delta(\tilde \xi_\mu^{a\star} \tilde \xi^{a\mu})}{\delta \theta} &=&   \xi^{\mu a\star}[L^a_\mu - 
[(D_\mu - \tilde D_\mu ) (\phi - \tilde \phi)^a + f^{a b c} (C^b - \tilde C^b)( \xi^c_\mu -\tilde \xi^c_\mu )]] - \kappa^{\mu a} \tilde \xi^a_\mu , \nonumber\\ 
\frac{\delta(\tilde {\bar\xi_\mu^{a\star}} \tilde {\bar \xi^{a\mu})}}{\delta \theta} &=& 
 \bar\xi_\mu^{a\star}[L^{\mu a} - h^{\mu a} +\tilde  h^{\mu a}] - \bar \kappa_{\mu a}  \tilde {\bar \xi^{a\mu}} , \nonumber\\
\frac{\delta(\tilde \phi^{a \star} \tilde \phi^a)}{\delta \theta} &=& - \tilde \phi^a \upsilon^a
- \phi^{a\star} [M^a -  f^{a b c} (C^b - \tilde C^b) (\phi^c - \tilde \phi^c ) ], \nonumber\\
\frac{\delta(\tilde {\bar \phi^{a\star}} \tilde {\bar \phi^a)}}{\delta \theta} &=& -\tilde {\bar \phi^a} \bar \upsilon^a 
- \bar \phi^{a\star} [\bar M^a -  \omega^a +\tilde  \omega^a], \nonumber\\
\frac{\delta(\tilde e^{a\star} \tilde e^a)}{\delta \theta} &=& -e^\star[N^a - (\lambda^a -\tilde\lambda^a)] 
-  \tilde e^a \tau^a.
 \label{q}
\end{eqnarray}
Adding all the equations of (\ref{q}) side by side, we get
\begin{eqnarray}
&& \frac{\delta}{\delta \theta} (\tilde A_\mu^{a\star} \tilde A^{a\mu} + \tilde {\bar 
\chi^{a\star}} \tilde \chi^a + \tilde {\bar \chi^a} \tilde \chi^{a\star} + \tilde b^{a\star} \tilde b^a + \tilde \xi_\mu^{a\star} \tilde \xi^{a\mu} 
+ \tilde {\bar \xi_\mu^{a\star}} \tilde {\bar \xi^{a\mu}} + \tilde \phi^{a\star} \tilde \phi^a + 
\tilde {\bar \phi^{a\star}} \tilde {\bar \phi^a} + \tilde e^{a\star} \tilde e^a)
 \nonumber\\&&= -\zeta^{a \mu} \tilde A^a_\mu - A_\mu^{a \star} [\psi^{a \mu} - 
(D^\mu - \tilde D^\mu ) (C - \tilde C)^a] - \bar \sigma^a  \tilde  C^a 
+ \bar C^{a \star}[\epsilon^a  - \frac{1}{2} f^{a b c}(C^b-\tilde{C}^b) (C^c -\tilde C^c)] \nonumber\\
& -&   \sigma^a  \tilde {\bar C^a} +  C^{a\star}[\bar \epsilon^a  - (b^a - \tilde b^a)] -  \upsilon^a 
\tilde \phi^a - \phi^{\star a} \left[ M^a - f^{a b c} (C^b - \tilde{C}^b )(\phi^c - \tilde{\phi}^c) \right] - \bar \upsilon^a \tilde {\bar \phi^a}
 \nonumber\\ &-&  \bar\phi^{\star a} (\bar M^a - \omega^a +\tilde \omega^a) - \tau^a \tilde e^a  -e^{\star a}[N^a - (\lambda^a - \tilde\lambda^a)] 
- \varpi^a \tilde b^a - b^{\star a} \chi^a - \kappa^{\mu a} \tilde \xi^a_\mu - 
\bar \kappa^{\mu a} \tilde {\bar \xi^a_\mu} \nonumber\\ 
&+&  
\xi^{\star \mu a} \left(L^a_\mu - \left[ (D_\mu - \tilde D_\mu ) (\phi - \tilde \phi )^a + f^{abc} (C^b  - \tilde C^b) (\xi^c_\mu - \tilde\xi^c_\mu ) \right] \right)  + \bar \xi^{\star \mu a} \left[\bar L^a_\mu - (h^a_\mu - \tilde{h}^a_\mu )\right],\label{mon}
\end{eqnarray}
which is nothing but the  gauge-fixed Lagrangian density
for shift symmetry $\tilde{{
\cal L}}_{gf+gh}$ given in (\ref{i}).
Now, one can  define the general  super-gauge-fixing fermion   in superspace  as 
follows
\begin{eqnarray}
\Phi(x, \theta) = \Psi (x)+ \theta (s\Psi),
\end{eqnarray}
which can further be expressed as
\begin{eqnarray}
\Phi(x, \theta) &=& \Psi (x)+ \theta \left[ - \frac{\delta \Psi}{\delta A^a_\mu}\psi^a _\mu + \frac{\delta \Psi}{\delta C^a} 
\epsilon^a
+ \frac{\delta \Psi}{\delta \bar C^a}\bar\epsilon^a - \frac{\delta \Psi}{\delta b^a} \chi^a
- \frac{\delta \Psi}{\delta \xi^a_\mu} L^a_\mu - \frac{\delta \Psi}{\delta \bar \xi^a_\mu} \bar L^a_\mu
- \frac{\delta \Psi}{\delta \phi^a} M^a\right. \nonumber \\ &-& \left.\frac{\delta \Psi}{\delta \bar \phi^a} \bar M^a - 
\frac{\delta \Psi}{\delta e^a} N^a \right ].
\end{eqnarray}
From this, the original gauge-fixing Lagrangian density   can be defined as the left 
derivation of super-gauge-fixing fermion with respect to $\theta$ 
as $\left[ \frac{\delta \Phi(x, \theta) }{\delta \theta}\right]$.

Hence, the complete effective action for the BF model in general gauge in the superspace is now 
given by
\begin{eqnarray}
{\cal L}_{eff}  &=& {\cal L}_0 +  \frac{\delta}{\delta \theta} \left[  \tilde A_\mu^{a\star} \tilde A^{a\mu} + \tilde {\bar 
\chi^{a\star}} \tilde \chi^a + \tilde {\bar \chi^a} \tilde \chi^{a\star} + \tilde b^{a\star} \tilde b^a + \tilde \chi_\mu^{a\star} \tilde \chi^{a\mu }
+ \tilde {\bar \chi_\mu^{a\star}} \tilde {\bar \chi^{a\mu}} + \tilde \phi^{a\star} \tilde \phi^a + 
\tilde {\bar \phi^{a\star}} \tilde {\bar \phi^a}+\tilde e^{a\star} \tilde e^a \right.
\nonumber\\
& +&\left. \Phi \right ].
\end{eqnarray}
Next, we will study the extended anti-BRST symmetry for BF model.

\section{Extended Anti-BRST Lagrangian Density}
In this section, we construct
the extended anti-BRST transformation under which the shifted Lagrangian density
for BF model remains invariant as follows,
\begin{eqnarray}
\bar s A^a_\mu &=&  A_\mu^{a\star} + (D_\mu - \tilde D_\mu )(\bar C - \tilde {\bar C})^a, \ \ 
\bar s \tilde A^a_\mu =  A_\mu^{a\star}, \ \ 
\bar s C^a = C^{a\star} - \frac{1}{2} f^{a b c} (C^b - \tilde C^b) (\xi^c_\mu - \tilde \xi^c_\mu ) , \nonumber\\ 
\bar s \tilde C^a &=& C^{a\star}, \ \ 
\bar s \bar C^a =\bar C^{a\star} - (b^a - \tilde b^a)   , \ \ 
\bar s \tilde {\bar C^a} = \bar {C^{a\star}}   , \ \ \bar s b^a = b^{a\star} + \chi^a,\ \
\bar s \tilde b^a = b^{a\star},  \nonumber\\ 
 \bar s \xi^a_\mu &=&  \xi_\mu^{a\star} - \left[(D_\mu - \tilde D_\mu ) (\phi^a - \tilde \phi^a )+ f^{a b c} ( C^b - \tilde C^b) (\xi^c_\mu - \tilde \xi^c_\mu )  \right] , \ \
\bar s \tilde \xi^a_\mu = \xi_\mu^{a\star} , \nonumber\\
\bar s \bar \xi^a_\mu &=&  \bar \xi_\mu^{a\star} - h^a_\mu +\tilde h^a_\mu, \ \
\bar s \tilde {\bar \xi^a_\mu} = \bar \xi_\mu^{a\star}  , \ \
\bar s \phi^a =  \phi^{a\star}  - f^{a b c} (C^b - \tilde C^b)(\phi^c - \tilde \phi^c )  , \ \
\bar s \tilde \phi^a =  \phi^{a\star}  , \nonumber\\
\bar s \bar \phi^a &=&  \bar \phi^{a\star} - \omega^a +\tilde\omega^a , \ \
\bar s \tilde {\bar \phi^a} =  \bar \phi^{a\star}  , \ \
\bar s e^a = e^{a\star} - (\lambda^a -\tilde \lambda^a), \ \
\bar s \tilde e^a = e^{a\star}.
\end{eqnarray}
The ghost fields associated with the shift symmetry  transform  under extended anti-BRST symmetry  as
\begin{eqnarray}
\bar s \psi_\mu^a  &=& \zeta_\mu^a , \ \ 
\bar s \epsilon^a  = \sigma^a , \ \ 
\bar s \bar \epsilon^a =  \bar \sigma^a  , \nonumber\\ 
\bar s \chi^a &=&   \varpi^a , \ \
\bar s L^a_\mu =  \kappa_\mu^a  , \ \ 
\bar s \bar L^a_\mu = \bar \kappa_\mu^a  , \nonumber\\
\bar s M^a &=&  \upsilon^a  , \ \ 
\bar s \bar M^a = \bar \upsilon^a  ,  \ 
\bar s N^a =  \tau^a.                        
\end{eqnarray}
From the nilpotency of above transformations demands that  the   auxiliary and antighost fields 
associated with the shift symmetry  transform as
\begin{eqnarray}
&&\bar s \zeta^a_\mu = 0   ,\ \ \  \bar s A_\mu^{a\star} = 0 ,\ \      
\bar s \sigma^a = 0  ,\ \  \ \bar s  C^{a\star} = 0  ,  \nonumber\\ 
&&\bar s \bar\sigma^a = 0 ,\ \  \ \bar s \bar C^{a\star} = 0,\ \ 
\bar s  \varpi^a = 0,\ \ \ \bar s b^{a\star} = 0 ,\nonumber\\
&&\bar s \kappa_\mu^a = 0   ,\ \ \  \bar s \xi_\mu^{a\star} = 0 ,\ \ 
\bar s \bar \kappa_\mu^a = 0   ,\ \ \  \bar s \bar \xi_\mu^{a\star} = 0 ,\nonumber\\
&&\bar s \upsilon^a = 0   ,\ \ \  \bar s \phi^{a\star} = 0 ,\ \ 
\bar s \bar \upsilon^a = 0   ,\ \ \  \bar s \bar \phi^{a\star} = 0 ,\nonumber\\ 
&&\bar s  \tau^a = 0   ,\ \ \  \bar s  e^{a\star} = 0.     
\end{eqnarray}
 
The gauge-fixing and ghost parts of the effective Lagrangian density are anti-BRST-exact also so it can be expressed as the anti-BRST variation of  this gauge-fixing fermion ($\bar \Psi$).

\section{Extended BRST and anti-BRST invariant superspace}
The extended BRST and anti-BRST invariant Lagrangian density  for BF model can be written in 
superspace with the help of two additional Grassmannian coordinates $\theta$ and $\bar{\theta}$. 
Requiring the field strength 
to vanish along unphysical directions $\theta$ and $\bar{\theta}$
direction,  we obtain the following  superfields:
\begin{eqnarray}
\textbf{A}^a_\mu (x,\theta , \bar \theta) &=& A^a_\mu (x) + \theta \psi^a_ \mu + \bar \theta[ A_\mu^{a\star} + (D_\mu - \tilde D_\mu )(\bar C - \tilde {\bar C} )^a]  + \theta \bar \theta \zeta^a_\mu , \nonumber\\
\tilde \textbf{A}^a_\mu (x,\theta , \bar \theta) &=& \tilde A^a_\mu (x) + \theta [\psi^a_ \mu - (D_\mu - \tilde D_\mu ) (C-\tilde C)^a ] + \bar \theta A_\mu^{a\star} + \theta \bar \theta \zeta^a_\mu , \nonumber\\
\textbf{C}^a(x,\theta , \bar \theta) &=& C^a (x)+ \theta \epsilon^a + \bar \theta[C^{a\star} - \frac{1}{2} f^{a b c} (C^b - \tilde C^b) (C^c - \tilde C^c )] + \theta \bar \theta\sigma^a, 
\nonumber\\
\tilde \textbf{C}^a(x,\theta , \bar \theta) &=& \tilde C^a(x) + \theta[\epsilon^a - \frac{1}{2} f^{a b c } (C^b - \tilde C^b) (C^c - \tilde C^c ) ] 
+ \bar \theta C^{\star a} + \theta \bar \theta \sigma^a, 
\nonumber\\
\bar\textbf{C}^a(x,\theta , \bar \theta) &=&  \bar C^a(x) + \theta \bar \epsilon^a + \bar \theta[\bar C^{\star a} - ( b - \tilde b)^a ] + \theta \bar \theta \bar \sigma^a, 
\nonumber\\
\tilde {\bar \textbf{C}^a}(x,\theta, \bar \theta) &=& \tilde {\bar C^a}(x) + \theta[\bar \epsilon^a - (b - \tilde b)^a] + \bar \theta \bar C^{\star a} + \theta \bar \theta \bar \sigma^a, \nonumber\\
\textbf{b}^a (x,\theta , \bar \theta) &=& b^a (x)+ \theta \chi^a + \bar \theta (b^{\star a} + \chi^a ) + \theta \bar \theta \varpi^a,
 \nonumber\\
\tilde \textbf{b}^a (x,\theta , \bar \theta) &=& \tilde b^a (x) + \theta \chi^a + \bar \theta b^{\star a} + \theta \bar \theta \varpi^a ,
 \nonumber \\
 {\xi}_\mu^a (x,\theta, \bar \theta) &=& \xi_\mu^a (x) + \theta L^a_\mu + \bar \theta\left( \xi_\mu^{a\star} - \left[(D_\mu - \tilde D_\mu ) (\phi^a - \tilde \phi^a )+ f^{a b c} ( C^b - \tilde C^b) (\xi^c_\mu - \tilde \xi^c_\mu )  \right] \right)+ \theta \bar \theta \kappa_\mu^{a}   , \nonumber\\
\tilde{ \bf{\xi}}_\mu^a (x,\theta , \bar \theta) &=& \tilde \xi_\mu^a (x) + \theta [L_\mu^{a} - ( D_\mu - \tilde D_\mu ) (\phi - \tilde \phi )^a + 
f^{a b c} (C^b - \tilde C^b) (\xi^c_\mu - \tilde \xi^c_\mu )] + \bar \theta \xi_\mu^{a\star} + \theta \bar \theta \kappa_\mu^a   , \nonumber\\
\bar {\bf{\xi}}_\mu^a (x,\theta , \bar \theta) &=& \bar \xi_\mu^a (x) + \theta \bar L^a_\mu + \bar \theta (\bar \xi_\mu^{\star a}- h^a_\mu +\tilde h^a_\mu) + \theta \bar \theta \bar \kappa_\mu^a   
, \nonumber\\
\tilde {\bar {\bf{\xi}}_\mu^a} (x,\theta, \bar \theta) &=& \tilde {\bar \xi_\mu^a }(x) + \theta (\bar L^a_\mu - h^a_\mu +\tilde h^a_\mu) + \bar \theta \bar \xi_\mu^{\star a} + \theta \bar \theta \bar \kappa_\mu^a   , \nonumber\\
 {\phi}^a (x,\theta , \bar \theta) &=& \phi^a (x)+ \theta M^a + \bar \theta \left(\phi^{a\star}  - f^{a b c} (C^b - \tilde C^b)(\phi^c - \tilde \phi^c ) \right) +  \theta \bar \theta \upsilon^a, 
\nonumber\\
\tilde {{\phi}}^a (x,\theta , \bar \theta) &=& \tilde \phi^a (x) + \theta ( M^a - f^{a b c} (C^b - \tilde C^b) (\phi^c - \tilde \phi^c )) + \bar \theta \phi^{\star a}  + \theta \bar \theta \upsilon^a,
 \nonumber\\
\bar{\bf{ \phi}}^a (x,\theta , \bar \theta) &=& \bar \phi^a (x)+ \theta \bar M^a + \bar \theta (\bar \phi^{\star a} - \omega^a+\tilde \omega^a)  + \theta \bar \theta \bar \upsilon^a, 
\nonumber\\
\tilde {\bar {\bf{\phi}}^a} (x,\theta , \bar \theta) &=& \tilde {\bar \phi^a} (x)+ \theta (\bar M^a - \omega^a +\tilde\omega^a ) + \bar \theta \bar \phi^{\star a} + \theta \bar \theta \bar \upsilon^a, \nonumber\\
\textbf{e}^a (x,\theta , \bar \theta) &=& e^a (x)+ \theta N^a + \bar \theta [e^{\star a} - (\lambda^a -\tilde \lambda^a )] + \theta \bar \theta \tau^a, 
\nonumber\\
\tilde \textbf{e}^a (x,\theta , \bar \theta) &=& \tilde e^a (x)+ \theta [N^a - (\lambda^a - \tilde \lambda^a)] + \bar \theta e^{\star a}  + \theta \bar \theta 
\tau^a.
\end{eqnarray}
With these expressions of superfields, we can calculate
\begin{eqnarray}
&-&\frac { 1}{2}  \frac {\partial}{\partial \bar \theta} \frac {\partial}{\partial 
\theta} \left(\tilde {\textbf{A}}^a_\mu \tilde {\textbf{A}}^{\mu a} +
 \tilde \chi^a  \tilde {\bar \chi^a} 
+ \tilde {\textbf{b}}^a  \tilde {\textbf{b}}^a + \tilde \xi^a_\mu  \tilde \xi^{\mu a}
+ \tilde {\bar \xi^a_\mu}  \tilde {\bar \xi^{\mu a}} + 
\tilde \phi^a \tilde \phi^a 
+ \tilde {\bar \phi^a} \tilde {\bar \phi^a}
+ \tilde {\textbf{e}}^a \tilde {\textbf{e}}^a \right)  \nonumber\\
  &=& 
 -\zeta^{a \mu} \tilde A^a_\mu - A_\mu^{a \star} [\psi^{a \mu} - 
(D^\mu - \tilde D^\mu ) (C - \tilde C)^a] - \bar \sigma^a  \tilde  C^a 
+ \bar C^{a \star}[\epsilon^a  - \frac{1}{2} f^{a b c}(C^b-\tilde{C}^b) (C^c -\tilde C^c)] \nonumber\\
& -&   \sigma^a  \tilde {\bar C^a} +  C^{a\star}[\bar \epsilon^a  - (b^a - \tilde b^a)] -  \upsilon^a 
\tilde \phi^a - \phi^{\star a} \left[ M^a - f^{a b c} (C^b - \tilde{C}^b )(\phi^c - \tilde{\phi}^c) \right] - \bar \upsilon^a \tilde {\bar \phi^a}
 \nonumber\\ &-&  \bar\phi^{\star a} (\bar M^a - \omega^a +\tilde \omega^a) - \tau^a \tilde e^a  -e^\star[N^a - (\lambda^a - \tilde\lambda^a)] 
- \varpi^a \tilde b^a - b^{\star a} \chi^a - \kappa^{\mu a} \tilde \xi^a_\mu - 
\bar \kappa^{\mu a} \tilde {\bar \xi^a_\mu} \nonumber\\ 
&+&  
\xi^{\star \mu a} \left(L^a_\mu - \left[ (D_\mu - \tilde D_\mu ) (\phi - \tilde \phi )^a + f^{abc} (C^b  - \tilde C^b) (\xi^c_\mu - \tilde\xi^c_\mu ) \right] \right)  + \bar \xi^{\star \mu a} \left[\bar L^a_\mu - (h^a_\mu - \tilde{h}^a_\mu )\right], 
\end{eqnarray}
which is nothing but the gauge-fixed Lagrangian density for shift symmetry.
Being the $\theta \bar \theta$ component of a super field,
 this Lagrangian density is manifestly invariant under both the extended BRST and the 
anti-BRST transformations. 

Now, we define the general super-gauge-fixing fermion   in superspace as
\begin{eqnarray}
 \Phi(x,\theta , \bar \theta) &=& \Psi(x) + \theta (s \Psi)
 + \bar \theta (\bar s \Psi) + 
 \theta \bar \theta (s \bar s \Psi),
\end{eqnarray}
which yields the original gauge-fixing and ghost part of the effective
effective Lagrangian density upon differentiation as follows,  $\mbox{Tr}\left[\frac {\partial}{\partial \theta}\left[\delta(\bar 
\theta ) \Phi(x,\theta , \bar \theta)\right]\right]$.

Therefore, the gauge-fixed Lagrangian density corresponding to BRST and shift symmetries for
BF model  can now be given as
\begin{eqnarray}
{\cal L}_{gf+gh}+\tilde{\cal L}_{gf+gh} 
 &=& -\frac { 1}{2}  \frac {\partial}{\partial \bar \theta} \frac {\partial}{\partial 
\theta} \left(\tilde {\textbf{A}}^a_\mu \tilde {\textbf{A}}^{\mu a} +
 \tilde \chi^a  \tilde {\bar \chi^a} 
+ \tilde {\textbf{b}}^a  \tilde {\textbf{b}}^a + \tilde \xi^a_\mu  \tilde \xi^{\mu a}
+ \tilde {\bar \xi^a_\mu}  \tilde {\bar \xi^{\mu a}} + 
\tilde \phi^a \tilde \phi^a 
+ \tilde {\bar \phi^a} \tilde {\bar \phi^a}
+ \tilde {\textbf{e}}^a \tilde {\textbf{e}}^a \right)  \nonumber\\ &+& \frac {\partial}{\partial \theta}
 \left[s(\bar \theta ) \Phi(x,\theta , \bar \theta)\right].
\end{eqnarray}
Therefore, we see that the BF model in superspace can be expressed in an elegant manner.
\section{Conclusion}
The $(3+1)$ dimensional BF model is subject of great interest due to its topological nature 
 and its some intriguing properties.   
In present work, we have considered $(3+1)$ dimensional BF model in Landau gauge and 
then we have shifted the Lagrangian to obtain the extended BRST and anti-BRST 
invariant (including some shift symmetry) BF model in BV formulation. 
The antifields corresponding to each field naturally arises.   Further  
we have provide the superfield description of BF model in superspace, where we show that the BV 
action for BF model can be written in a manifestly extended BRST invariant manner in a 
superspace by considering one additional Grassmann (fermionic) coordinate. 
However,  we need two additional Grassmann coordinates to express  both the extended BRST 
and extended anti-BRST invariant BV actions of BF model in superspace.\\\\\\ 
\textbf{{Conflicts of Interest}}\\\\
The author  declares that there are no conflicts of interest
regarding the publication of this paper.

\begin{acknowledgements}
The author is grateful to Dr. Sudhaker Upadhyay for his suggestions in preparation to manuscript. 
 \end{acknowledgements}

\end{document}